\documentclass[aps,prl,showpacs,preprintnumbers,amsmath,amssymb]{revtex4}
\usepackage{graphicx}
\usepackage{epstopdf}
\begin{document}
\begin{abstract}
 In this work we experimentally demonstrate for the first time  a  recently proposed criterion addressed to detect nonclassical behavior in the fluorescence emission of ensembles of single-photon emitters. In particular, we apply the method to study clusters of NV centres in diamond observed via single-photon-sensitive confocal microscopy. Theoretical considerations on the behavior of the parameter at any arbitrary order in presence of poissonian noise are presented and, finally, the opportunity of detecting manifold coincidences is discussed.
\end{abstract}

\title{Direct experimental observation of nonclassicality in ensembles of single photon emitters}

\author{E. Moreva$^1$, P. Traina$^1$, J. Forneris$^{2,3}$,  I. P. Degiovanni$^1$, S. Ditalia Tchernij$^3$, F. Picollo$^3$, G. Brida$^1$, P. Olivero$^{3,2}$and M. Genovese$^{1,2}$.}
\affiliation{$^1$Istituto Nazionale di Ricerca Metrologica, Strada delle cacce 91, Turin, Italy}
\affiliation{$^2$Istituto Nazionale di Fisica Nucleare (INFN) Sez. Torino, Torino, Italy}
\affiliation{$^3$Physics Department and NIS Centre of Excellence - University of Torino, Torino, Italy}



\newcommand{\ket}[1]{\mbox{\ensuremath{|#1\rangle}}}
\newcommand{\bra}[1]{\mbox{\ensuremath{\langle#1|}}}


\flushbottom
\maketitle

\thispagestyle{empty}

\section*{Introduction}

One of the most debated issues in quantum mechanics is related to understanding the boundary separating the counterintuitive behavior of the systems governed by the quantum laws from the classical, familiar properties of the macroscopical systems. This transition also manifests itself in the realm of optics \cite{bondani} where, even if it is undoubtful that the radiation emitted by any possible source of light is indeed composed by an ensemble of individual photons, the properties of classical sources differ consistently from those of non-classical ones, in particular single photon sources (SPS), that have found many experimental and reliable realization in systems such as heralded sources based on parametric down-conversion \cite{sramelow,sorgentina,krapick,fortsch,montaut,oxborrow,eiseman}, quantum dots, trapped ions, molecules and colour centres in diamond\cite{dia1,dia2,dia3,dia4,dia5,dia6}.
Since non-classical optical states have now become a fundamental resource for quantum technology, the determination of nonclassicality for a state is not only important for studies concerning boundaries form quantum to classical world, but also represents an important tool for quantifying such resources.
 There exists a huge literature on the characterization of SPSs \cite{sps}. Most of the techniques rely on the sampling of the second order autocorrelation function (or Glauber function) 
 \begin{equation}
 g^{(2)}(\tau=0)=\frac{\langle I(t)I(t+\tau)\rangle}{\langle I(t)\rangle\langle I(t+\tau)\rangle}\rvert_{\tau=0},
 \end{equation}
 
  whose value is never smaller than $1$ for classical light, while it is lower than $1$ for sub-poissonian light,  and in particular vanishes for single photon states, where $g^{(2)}(0)=0$ is 
 expected in the ideal case. This quantity has been shown to be substantially equivalent to the parameter $\alpha^{(2)}$ introduced by Grangier et al.\cite{grangier} (and throughout the paper we will refer to this parameter as $g^{(2)}$ without distinction), which is experimentally measured as the ratio between  the coincidence probability at the ouput of a Hanbury-Brouwn and Twiss interferometer (HBT)\cite{hbt}, basically a $50:50$ beamsplitter connected to two non-photon-number-resolving (non-PNR) detectors, and the product of the click probabilities at the two detectors. This parameter can be generalized to account for statistical properties of $N$-fold coincidence events at the outputs of detector-tree apparata and several techniques for the reconstruction of optical states as well as quantum enhanced imaging techniques are allowed by the experimental sampling of $g^{(N)}$ functions\cite{elizabeth,sres,vonz1,vonz2}. Unfortunately, the amount of background light can affect the measurement, leading to a camouflage of the quantum characteristics due to noise. More specifically, in practical cases, when sampling $g^{(2)}(0)$  to characterize single emitters one cannot distinguish between the true quantum signal and background light contribution and, in extreme cases, one is not able to detect a single emitter drowned in dominant noise bath. Recently a novel criterion able to reveal non classical light form large numbers of independent SPS has been proposed\cite{Filip}. According to the theoretical predictions, an experimental implementation of this criterion would be extremely advantageous not only because it would allow to spot non-classical signatures in the emission of clusters of emitters, but also because it can be shown that this technique is extremely robust in the presence of poissonian noise, the parameter under test being absolutely independent from this kind of noise contribution (even if dominant).

In this work we experimentally apply for the first time the criterion\cite{Filip} to directly detect non-classical  emission from ensembles of SPSs based on Nitrogen-Vacancy (NV) centers in nano-diamond observed by means of a confocal microscope coupled to four non-PNR single-photon detectors in a detector-tree configuration.

\section*{Results}

\subsection*{Theoretical model}
In general, the system considered here is an ensemble of $M$ single-photon-emitters, each coupled by the detection system with an efficiency $\eta_{\alpha}$ ($\alpha=1,...,M$) detected by $N$ non-PNR detectors connected  by a generalized $N$-dimensional beam-splitter (BS). Each detection channel has an overall efficiency (due to BS unbalance and detector efficiency) $\xi_{i}$ ($i=1,...,N$). 

The generalized $g^{(N)}(0)$ function is expressed in terms of detection probabilities as:
\begin{equation}
g^{(N)}(0)=\frac{P_{click^{\otimes N}}}{\prod_{i=1}^N P_{click[i]}}\label{eq:g}
\end{equation}
where $P_{click^{\otimes N}}$ is the probability of $N$-fold coincidence at the output of the detector-tree, and $P_{click[i]}$ is the probability for the $i$-th detector to fire. As stated above, the condition $g^{(N)}(0)=1$ can be used to discriminate between classical and nonclassical states, while ideally $g^{(N)}(0)=0$ for any order of $N$ for single-photon states.
Instead, the nonclassicality criterion  under study\cite{Filip} is  expressed  by the fact that for any classical system  the following proposition is verified:
\begin{equation}
\theta^{(N)}(0)=\frac{P_{0^{\otimes N}}}{\prod_{i=1}^N P_{0[i]}}>1,
\label{eq:criterion}
\end{equation}
where, $P_{0[i]}$ is the \emph{no-click} probability at the $i$-th detector and $P_{0^{\otimes N}}$ is the probability that all the $N$ detectors of the detector-tree do not click in correspondance of an excitation event.
In the methods section, the derivation of the expression for $\theta^{(N)}$ and $g^{(N)}$ functions in presence of poissonian noise is presented. 

\subsection*{Experimental data}

In the following we will describe the results on the implementation of the nonclassicality criterion \cite{Filip} by the characterization of three fluorescent objects in a nano-diamonds sample both in terms of $g^{(2)}$ and  $\theta^{(2)}$ functions. To perform this study, the single-photon-sensitive confocal microscope was coupled to a detector-tree configuration of four detectors registering 6 different two-fold coincidences in a 40-ns temporal window. The width of the window was chosen to be compatible with the lifetime of the centers (around 25 ns). The objects under study are dubbed Item-1, Item-2, Item-3. Item-1 is reasonably compatible with a single photon emitter having $g^{(2)}$ value below $0.5$ ($g^{(2)}_{I_1}(0) = 0.407 \pm 0.012$) 
if no artificial noise is added, while Item-2, Item-3 are clusters of unknown quantities of single photon emitters (respectively $g^{(2)}_{I_2}(0) = 0.832 \pm 0.004$ and $g^{(2)}_{I_3}(0) = 0.66 \pm 0.01$, always without 
noise).
In order to simulate poissonian noise, a power-regulable laser source at a wavelength falling in the detection spectral window was reflected directly in the coupling pinhole of the microscope. To analyze the robustness of the parameter with respect to noise, every measurement was repeated for three different values of intensity of the noise source, measured as the count rate due only to the poissonian source (excitation light off) at the single channel (1- noise off, 2-
$10000$ counts/s,3- $25000$ counts/s).
\begin{figure}[ht]
\centering
\includegraphics[scale=0.5]{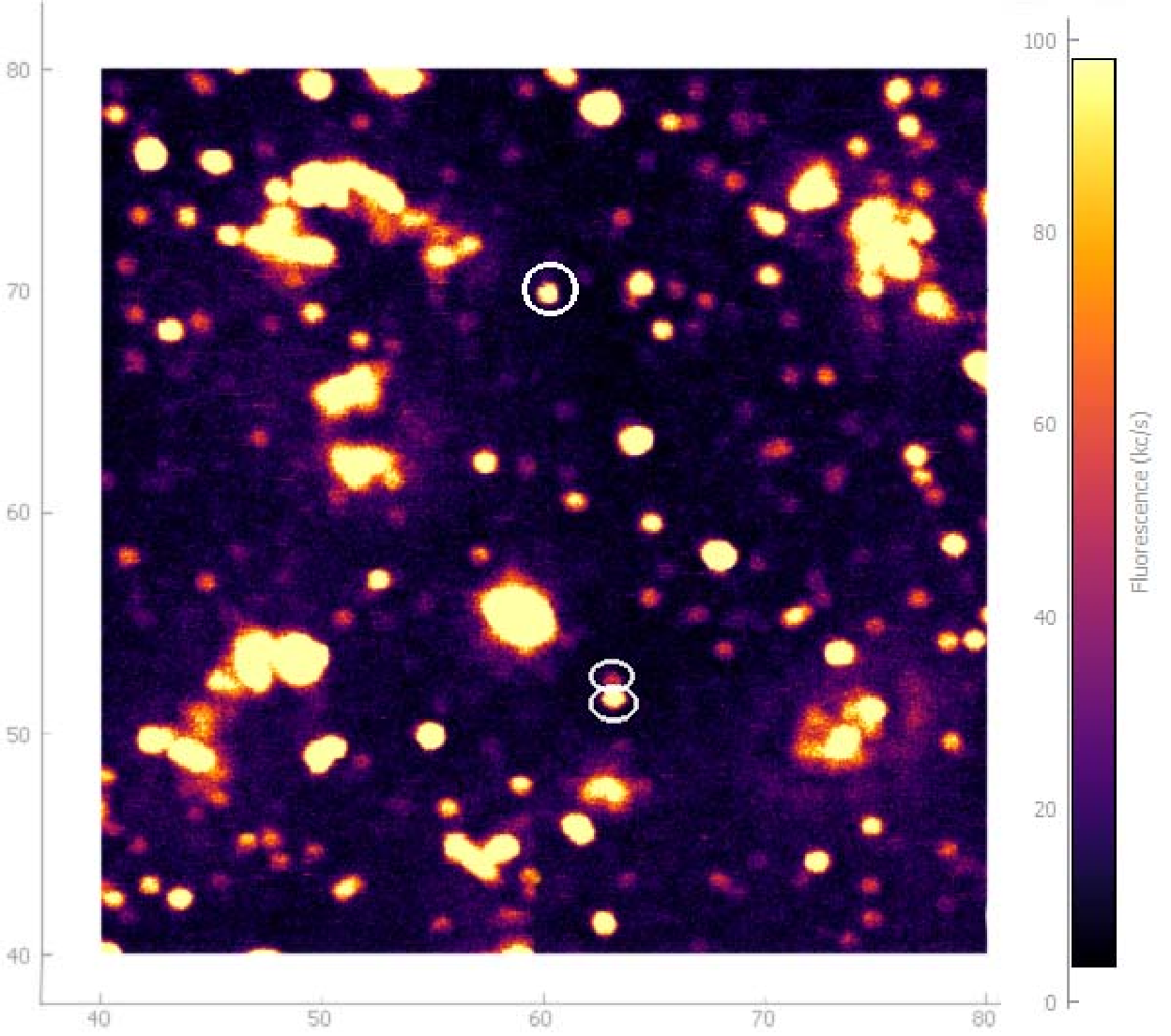} 
\caption{Typical fluorescence map of the selected area of the sample obtained with the single-photon-sensitive confocal microscope. The three highlighted spots correspond, respectively from top to bottom, to Item-2, Item-1 and Item-3. The acquisition software used is "Qudi", developed at Ulm University\cite{qudi}}
\label{fig:map}
\end{figure}

  For testing the stability of the experimental system, the measurement without poissonian noise has been repeated after the measurement including noise and the $g$ and $\theta$ parameters are found to be consistent with the first measurements. As an example, this measurement in the case of Item-1 is shown as a red dot. Also, to test the capability of our setup to detect nonclassical behavior, we performed the measurement of $\theta^{(2)}$ and $g^{(2)}$ parameters on the light reflected by a non-fluorescent nano-diamond present in the sample. This kind of object appears not distinguishable from the emitters in a confocal map but does not produce antibunching (being produced by coherent light) and can be recongnized only by spectral characteristics. As expected, this object showed  clear signatures of classical emission (($1-g^{(2)}(0) = 0.004 \pm 0.005$, $1-\theta^{(2)}(0) = (-4 \pm 2)*10^{-8}$ ).

\begin{figure}[ht]
\centering
\includegraphics{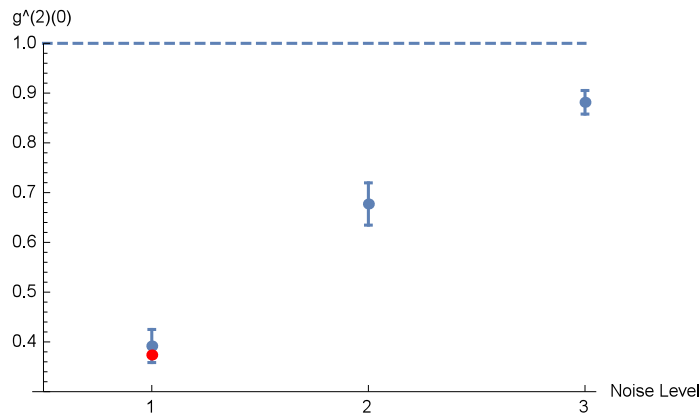}
\caption{Plot of $g_{I_1}^{(2)}(0)$ measured for three different levels of poissonian noise (1- noise off, 2- $10000$ counts/s due to noise, 3-$25000$ counts/s due to noise). Each measure is the average of 6 runs, registered in parallel in every two-fold combination of the detector-tree branches, of $200$ s. Excitation rate is $5$ MHz. The red dot corresponds to a successive repetition of the first measurement to test the stabilty of the apparatus.}
\label{fig:singleg2}
\end{figure}

\begin{figure}[ht]
\centering
\includegraphics{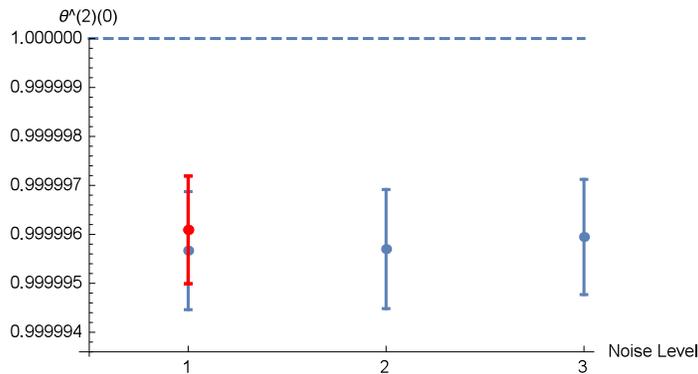}\\
\caption{Plot of $\theta_{I_1}^{(2)}(0)$ measured  for three different levels of poissonian noise (1- noise off, 2- $10000$ counts/s due to noise, 3- $25000$ counts/s due to noise). Same measurement settings as Figure \ref{fig:singleg2}. The red dot corresponds to a successive repetition of the first measurement to test the stabilty of the apparatus.}
\label{fig:singleth2}
\end{figure}

\begin{figure}[ht]
\centering
\includegraphics{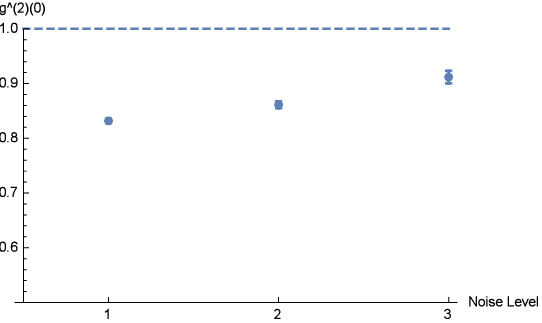} 
\caption{Plot of $g_{I_2}^{(2)}(0)$ measured  for three different levels of poissonian noise (1- noise off, 2- $10000$ counts/s due to noise,3- $25000$ counts/s due to noise). Same measurement settings as Figure \ref{fig:singleg2}.}
\label{fig:cluster2g2}
\end{figure}

\begin{figure}[ht]
\centering
\includegraphics{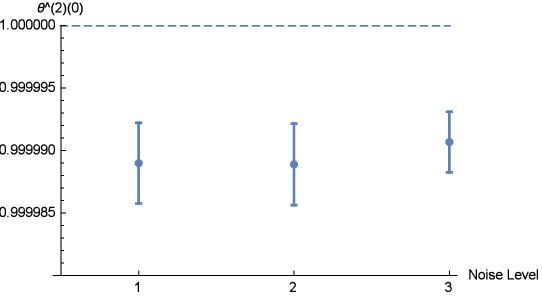}
\caption{Plot of $\theta_{I_2}^{(2)}(0)$ measured  for three different levels of poissonian noise (1- noise off, 2- $10000$ counts/s due to noise,3- $25000$ counts/s due to noise ). Same measurement settings as Figure \ref{fig:singleg2}.}
\label{fig:cluster2th2}
\end{figure}

\begin{figure}[ht]
\centering
\includegraphics{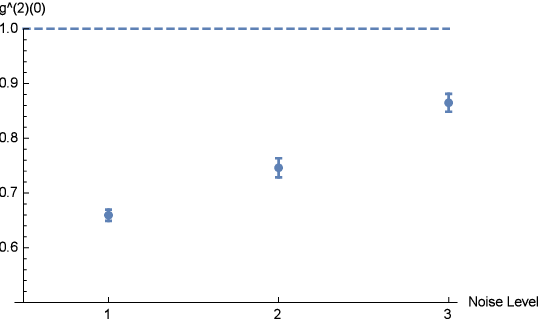} 
\caption{Plot of $g_{I_3}^{(2)}(0)$ measured  for three different levels of poissonian noise (1- noise off, 2- $10000$ counts/s due to noise, 3- $25000$ counts/s due to noise). Same measurement settings as Figure \ref{fig:singleg2}.}
\label{fig:cluster1g}
\end{figure}

\begin{figure}[ht]
\centering
\includegraphics{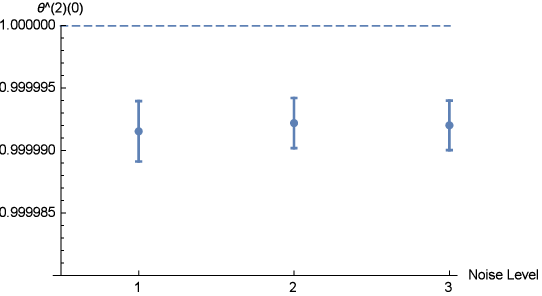}
\caption{Plot of $\theta_{I_3}^{(2)}(0)$ measured  for three different levels of poissonian noise (1- noise off, 2- $10000$ counts/s due to noise, 3- $25000$ counts/s due to noise). Same measurement settings as Figure \ref{fig:singleg2}.}
\label{fig:cluster1th}
\end{figure}

\begin{figure}[ht]
\centering
\includegraphics[scale=1.2]{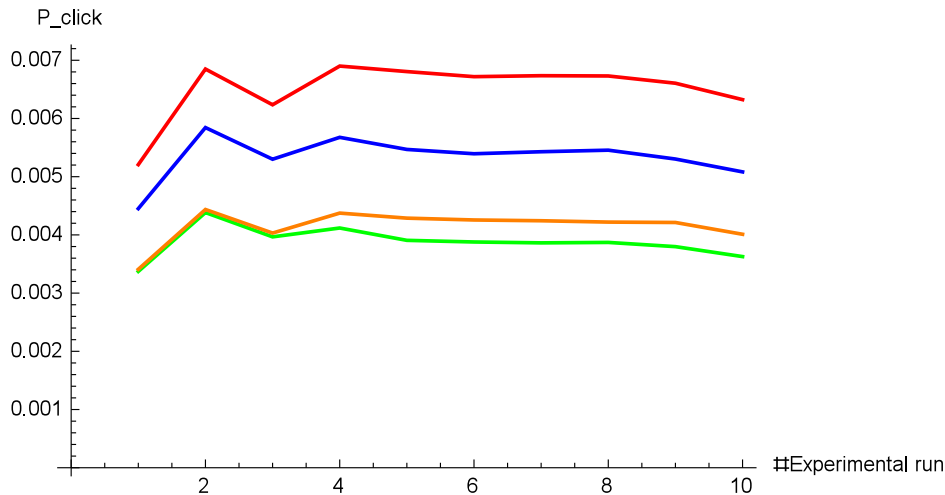}
\caption{Click-probabilities at the four outputs of the detector-tree while observing Item-3. Same measurement settings as Figure \ref{fig:singleg2}.}
\label{fig:cluster1pi}
\end{figure}


\begin{figure}[ht]
\centering
\includegraphics[scale=1.2]{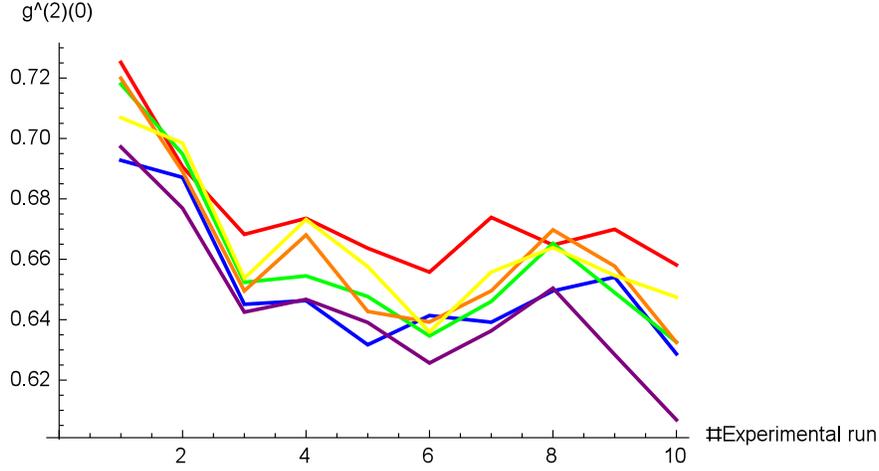}
\caption{Values of $g^{(2)}(0)$ independently measured for each two-fold combination among the four outputs of the detector-tree while observing Item-3. Same measurement settings as Figure \ref{fig:singleg2}.}
\label{fig:cluster1g2ij}
\end{figure}

\begin{figure}[ht]
\centering
\includegraphics[scale=1.2]{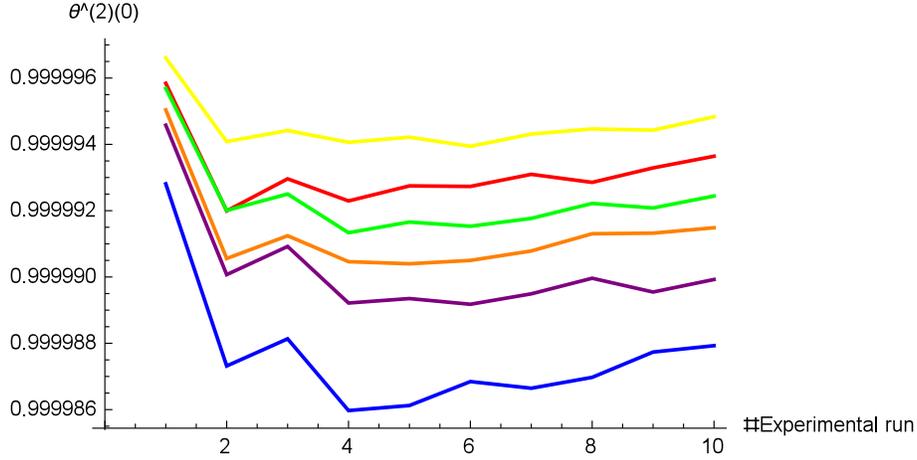}
\caption{Values of $\theta^{(2)}(0)$ independently measured for each two-fold combination among the four outputs of the detector-tree while observing Item-3. Same measurement settings as Figure \ref{fig:singleg2}. It is observed that, as opposed to the $g^{(2)}(0)$ case, $\theta^{(2)}(0)$ are less consistent due to the dependence of the parameter on the efficiency of the detection channel.}
\label{fig:cluster1th12}
\end{figure}






\section*{Discussion}

As predicted by the theoretical proposal, the experimental data clearly demonstrate the two advantages of the tested non-classicality criterion with respect to $g^{(2)}$ characterization: firstly, the parameter estimation is more robust against poissonian noise since the three $\theta^{(2)}$ values for each object are perfectly compatible, even if $g^{(2)}$ values differ considerably for different noise contributions; secondly, the deviation from classicality is even stronger when the object under study is not a single emitter but instead an ensemble of them. For this reason this method allows detecting non-classical behavior of quantum emitters without the necessity of isolating a single one. 
On the other hand, the $\theta$ function presents some disadvantages, since its value depends strongly on the efficiency of the channel
$\xi$ (including detection efficiency and splitting ratio of the detector-tree, see the methods section) as opposed to $g^{(2)}$ which is independent from the balance of the branches of the detector tree. This can be observed in the plots in Fig.s \ref{fig:cluster1pi}, 
\ref{fig:cluster1g2ij}, \ref{fig:cluster1th12}, where respectively, the click-probabilities of the single channels, 
the experimental $g^{(2)}$ and $\theta^{(2)}$ values in the characterization of Item-3 are shown. It results that the unbalance of the detector tree in terms of detection efficiency which is observable in Fig.s \ref{fig:cluster1pi}, 
is compensated when sampling $g^{(2)}$ (Fig. \ref{fig:cluster1g2ij}), delivering 6 fairly consistent independent estimates,  while the values of $\theta^{(2)}$ obtained for different pairs of channels are not consistent. It must be reminded anyway that the average of the latter values is below the classical limit in a $3\sigma$ confidence band, thus this inconsistency does not affect the observation of nonclassical behavior.

In our experiment we limited the analysis to the second order of correlation of $\theta$ and $g$ functions. Since the family of nonclassicality criteria under test is defined for any arbitrary order, it is reasonable to investigate on the possible advantages/disadvantages of experimentally sampling three- and four-fold coincidences as well. The calculations on the behavior of the $\theta^{(N)}$ functions clearly reveal the independence of the value of the parameter respect to the noise level for all the orders of correlation,
exacltly as we observed in the second order case and at variance with the behavior of $g$-functions (see methods section). Even if our detection apparatus is capable of detecting up to four-fold coincidences, for the three objects under study the three- and four-fold coincidence rates were extremely low (respectively $10^{-7}$ and $10^{-9}$, less than the associated statistical uncertainty for the brightest object, Item-2), so that no significant consideration on high order behavior could be extracted in our experimental conditions.

In conclusion, in this paper we have demonstrated experimentally the advantage of non-classicality criteria based on $\theta^(N)$ function proposed in\cite{Filip}. Our results pave the way both for studying quantum - classical boundary and for quantifying resources needed in quantum technologies.

\section*{Methods}

\subsection*{Sample preparation}

This work was performed on synthetic nanodiamond (ND) powders produced by
ElementSix$^{TM}$ by disaggregation of High Pressure High Temperature (HPHT) syntetic diamond with a nominal size of distribution comprised between $10$ nm and $250$ nm. The powders were classified as Ib type, with a nominal substitutional N concentration of 10 $\div$ 100 ppm, and contained by a low amount of native NV centers as a consequence of the high crystalline quality of the batch. NV centers were therefore fabricated through the introduction of radiation-induced vacancies and a subsequent thermal annealing \cite{Rondin}. The ND powders were firstly exposed to an acid bath ($H_2SO_4:HNO_3 = 9:1$ solution; 72 hours at $100^\circ C$) to remove organic contaminations and graphite, and subsequently dispersed over a suitable substrate for ion irradiation.Then,  they were irradiated with a $2$ MeV $H^+$ beam at the AN2000 Accelerator of the INFN National Laboratories of Legnaro (Italy).
An irradiation fluence of $5x10^{12}$ protons/cm$^{2}$ was chosen, based on the ND median size and the NV centers creation efficiency \cite{Pezzagna}, as the optimal condition to fabricate 150 nm sized NDs containing one individual NV center. Then, the powders were exposed to a thermal treatment ($800 ^\circ C$ for 1 h, in a 800 mbar controlled $N_2$ atmosphere) in order to promote the formation of NV centers.
After the annealing process, an additional cleaning step was performed by a 30 min sonic bath in $H_2SO_4$ , followed by a cleaning in Piranha solution ($H_2SO_4: H_2O_2 = 3:1$) to remove organical residuals and to dissolve metal oxides and carbonates contents from the ND powders.
The samples were finally dispersed on cover-slip glass substrates for a subsequent optical investigation.

\subsection*{Experimental setup}
The sample is observed via a single-photon-sensitive confocal microscope connected to a detector-tree configuration of 4 detectors. The excitation light is provided by a solid state laser at 532 nm (PICOQUANT LDH-PFA-530L) in pulsed regime (5 MHz repetition rate, 50 ps FWHM) whose output, coupled into a single mode fibre, is collimated by a $4\times$ objective.
A dichroic mirror (Long-pass at 570 nm)  reflects the excitation light (3 mW maximum) inside the oil immersion objective (Olympus, $100\times$, NA = 1.3
) focusing inside the sample and transmits the fluorescence light towards the detecting apparatus. The sample is mounted on a closed-loop XYZ piezo-electric stage, remotely controlled via PC, allowing submicrometric-resolution positioning in a $80\mu$m $\times 80 \mu$m area range. The fluorescence light (occurring within a  $640 - 800$ nm spectral window) is collected by the same objective used for excitation and then passes through a dichroic mirror and a long-pass filter in order to obtain a suitable attenuation ($>10^{12}$) of the pump component. Then, the signal is focussed by a $f = 100$ mm achromatic doublet and coupled to a $50$ $\mu$m multimode optical fibre.
The fiber leads to a detector-tree configuration realized by means of two integrated $50:50$ beam-splitters in cascade connecting to four Single Photon Avalanche Photo-diodes (Perkin-Elmer SPCM-AQR), operating in Geiger mode. This configuration, reproducing 6 parallel HBT interferometers \cite{hbt}, allows the detection of all the two-fold coincidences among the detection channels and to obtain six independent estimations of the second order  autocorrelation  functions ($g^{(2)}$). The signal counts and coincidences are measured via a Id Quantique ID800 time-to-digital converter.
The pulses (60 ps FWHM) of the laser simulating the poissonian noise (PICOQUANT LDH-D-C-690), emitting at $685$ nm, inside the detection window, was electronically synchronized with the excitation laser emission. This laser was directly coupled to the pinhole of the detection system.

\subsection*{Calculation of $\theta^{(N)}$ and $g^{(N)}$ in presence of poissonian noise}
Given $n$ incoming photons entering in the detector-tree, they are  distributed in the $N$ channels following the multinomial probability $\frac{n!}{\prod_{i=1}^N k_i!}\Bigl(\frac{1}{N}\Bigr)^n$ corresponding to $k_i$ photons in the i-th channel (satisfying $\sum_i k_i=n$).
In each channel of the detector tree, the probability of observing a \emph{no-click} event given $k_i$ photons is $(1-\xi_i)^{k_i}$, thus the $click$ probability is $1-(1-\xi_i)^{k_i}$. This derives from the POVM (positive operator-valued operators) of photodetection of the single-photon detector at the end of each channel of the detector tree:
\begin{equation}
\hat Q_{click}=\sum_{n=0}^{+\infty}[1-(1-\xi_i)^n]\ket{n}\bra{n}, \quad \hat Q_{noclick}=\sum_{n=0}^{+\infty}(1-\xi_i)^n\ket{n}\bra{n}
\end{equation}
Starting from this one can define the POVM of the single detection of the detector tree as
\begin{equation}
\hat Q_{[i]}^{[Single]}(0)=\sum_{n=0}^{+\infty} Q_{[i]}^{[Single]}(0|n)\ket{n}\bra{n }, \quad \hat Q_{[i]}^{[Single]}(1)=\hat I-\hat Q_{[i]}^{[Single]}(0)
\end{equation}
where $Q_{[i]}^{[Single]}(0|n)=(1-\xi_i/N)^n$ is the probability that $0$ out of $n$ incoming photons are detected per excitation pulse. Since the measurement is phase-insensitive, the operators have diagonal form in the Fock basis and, due to the nature of non-PNR detectors (able only to distinguish between dark and light) the possible outcomes are "0" (the detector does not click) and "1" (the detector clicks).

Analogously, one can obtain the POVM  associated to the no-click in all the outputs of the detector tree as
\begin{equation}
\hat Q^{[\otimes N]}(0)=\sum_{n=0}^{+\infty}Q^{[\otimes N]}(0|n)\ket{n}\bra{n}
\end{equation}
where
$Q^{[\otimes N]}(0|n)=(1-\frac{{\sum_{i=1}^N}\xi_i}{N})^n$.

Finally, thethe POVM  of $N$-fold coincidence 
results :
\begin{equation}\label{eq:nfold}
\hat Q^{[\otimes N]}(N)=\sum_{n=0}^{+\infty}Q^{[\otimes N]}(N|n)\ket{n}\bra{n},
\end{equation}
where $Q^{[\otimes N]}(N|n)$ has in general a rather complicated form, but under the hypotesis that the detection system is a tree of perfectly balanced identical detectors ($\xi_i=\xi, \forall i$), it reduces to:
\begin{equation}
Q^{[\otimes N]}(N|n)=\sum_{r=0}^N(-1)^r\frac{N!}{r!(N-r)!}(1-\frac{r \xi}{N}).
\end{equation}
It follows that the probability $P_{0^{\otimes N}}=tr[\hat \rho \hat Q^{[\otimes N]}(0)]$ ($P_{0[i]}=tr[\hat \rho \hat Q_{[i]}^{[Single]}(0)]=P_0$) in eq. \ref{eq:criterion}, $\hat \rho$ being the density matrix describing the quantum state of the ensemble of emitters,  can be expressed in the form:
\begin{equation}\label{eq:ptot}
\sum_{n=0}^\infty \sigma^n p_n,
\end{equation}
 where $p_n=\bra{n}\hat \rho\ket{n}$ is the probability distribution of the photons and $\sigma^n$ is equal to $(1-\xi)^n$ ($(1-\frac{\xi}{N})^n$).
We study the case of single emitters' fluorescence in presence of poissonian noise. The photon-number probability ditribution in this case is
\begin{equation}
p_n=\sum_{m=0}^M\sum_{k=0}^\infty\delta_{n,m+k}P_{sps}(m)P_{\lambda}(k),
\end{equation}

where, assuming that all the emitters in the ensemble are coupled with the same efficiency ($\eta_{\alpha}=\eta, \forall \alpha$), $P_{sps}(m)=\frac{M!}{m!(M-m)!}\eta^n(1-\eta)^{M-m}$ is the distribution of the photons of the emitters, $P_{\lambda}(k)=\frac{\lambda^k e^{-\lambda}}{k!}$ is the distibution of the poissonian light and $\delta_{x,y}$ is the kronecker delta.
Substituting in eq. \ref{eq:ptot} the suitables value for $\sigma$, one gets:
\begin{eqnarray}
P_{0^{\otimes N}}&=&(1-\eta\xi)^Me^{-\lambda\xi}\label{eq:ciao1}\\
P_{0[i]}&=& P_0=(1-\frac{\eta\xi}{N})^Me^{-\frac{\lambda\xi}{N}}.\label{eq:ciao2}
\end{eqnarray}
Finally, substituting eqs. \ref{eq:ciao1},\ref{eq:ciao2} in eq. \ref{eq:criterion}, the $\lambda$-dependant terms appear as equal factors both in the numerator and in the denominator of the ratio and are simplified, resulting:
\begin{equation}
\theta^{(N)}(0)=\frac{(1-\eta\xi)^M}{(1-\frac{\eta\xi}{N})^{MN}},
\end{equation}
Thus, under our assumptions, the parameter $\theta^{(N)}$ estimation is independent from the poissonian contribution at any order $N$ (at variance with $g^{(N)}$).

This parameter must be compared with $g^{(N)}$-function that is expressed according to eq. \ref{eq:g}. In order to calculate it in analogy with the expession of $\theta^{(N)}$, we must first of all write

the probability of $N$-fold coincidence:
\begin{equation}
P_{click^{\otimes N}}=tr[\hat \rho \hat Q^{[\otimes N]}]=\sum_{n=0}^\infty Q^{[\otimes N]}(N|n)p_n=\sum_{r=0}^N (-1)^r\frac{N!}{r!(N-r)!}\Bigl(1-\frac{\eta r\xi}{N}\Bigr)^ M e^{-\frac{\lambda r\xi}{N}},
\end{equation}
leading to
\begin{equation}
g^{(N)}(0)=
\frac{P_{click^{\otimes N}}}{(P_{click})^N}
=\frac{\sum_{r=0}^N (-1)^r\frac{N!}{r!(N-r)!}\Bigl(1-\frac{\eta r\xi}{N}\Bigr)^ M e^{-\frac{\lambda r\xi}{N}}}{{\Bigl[1-(1-\frac{{\eta\xi}}{N})^M e^{-\frac{\lambda\xi}{N}}\Bigr]^N}},
\end{equation}
where, in accordance with eq. \ref{eq:ciao2} we used $P(click)=1-P_0$.
It is clear that, in opposition with the $\theta^{(N)}$ case, the contribution of  the poissonian terms to $g^{(N)}$ cannot be eliminated.

\subsection*{Explicit $\theta^{(N)}$ and $g^{(N)}$ expressions}

The following are the explicit expressions of $\theta^{(2)}$,  $\theta^{(3)}$, $\theta^{(4)}$ as functions of the click and coincidence probabilities at the detectors:
\begin{eqnarray*}
\theta_{[ij]}^{(2)}&=&\frac{1-P_{click[i]}-P_{click[j]}+P_{click^{\otimes 2}[ij]}}{(1-P_{click[i]})(1-P_{click[j]})}\label{eq:theta1}\\
\theta_{[ijk]}^{(3)}&=&\frac{1-P_{click[i]}-P_{click[j]}-P_{click[k]}+P_{click^{\otimes 2}[ij]}+P_{click^{\otimes 2}[ik]}+P_{click^{\otimes 2}[jk]}-P_{click^{\otimes 3}[ijk]}}{(1-P_{click[i]})(1-P_{click[j]})(1-P_{click[k]})}\label{eq:theta2}\\
\theta_{[ijkl]}^{(4)}&=&\frac{1}{(1-P_{click[i]})(1-P_{click[j]})(1-P_{click[k]})(1-P_{click[l]})}( 1-P_{click[i]}-P_{click[j]}-P_{click[k]}-P_{click[l]}+\dots\\ &\dots&+P_{click^{\otimes 2}[ij]}+P_{click^{\otimes 2}[ik]}+P_{click^{\otimes 2}[il]}+P_{click^{\otimes 2}[jk]}+P_{click^{\otimes 2}[jl]}+P_{click^{\otimes 2}[kl]}+\dots\\ &\dots&-P_{click^{\otimes 3}[ijk]}-P_{click^{\otimes 3}[ijl]}-P_{click^{\otimes 3}[ikl]}-P_{click^{\otimes 3}[jkl]}+P_{click^{\otimes 4}[ijkl]}),
\label{eq:theta3}
\end{eqnarray*}
where, for instance $P_{click[i]}$ is the click probability at the $i$-th detector, $P_{click^{\otimes 2}[ij]}$ is the two-fold coincidence probability between channels $i$ and $j$ and $P_{click^{\otimes 3}[ijk]}$ ($P_{click^{\otimes 4}[ijkl]}$) is the three-(four-)fold coincidence probability among channels $i$,$j$,$k$ ($i$,$j$,$k$,$l$). The latter probabilities are experimentally sampled from single channel the detection ($N_i$), the two-($N_{ij}$),three-($N_{ijk}$) and four-fold ($N_{ijkl}$) coincidence rates respectively as  $P_{click[i]}=N_i/N_{TR}$, $P_{click^{\otimes 2}[ij]}=N_{ij}/N_{TR}$, $P_{click^{\otimes 3}[ijk]}=N_{ijk}/N_{TR}$, $P_{click^{\otimes 4}[ijkl]}=N_{ijkl}/N_{TR}$ and $N_{TR}$ is the rate of excitation events (the repetition rate of the excitation laser).
Analogously, the $g$-functions are estimated as:
\begin{eqnarray}
g_{[ij]}^{(2)}&=&\frac{P_{click^{\otimes 2}[ij]}}{P_{click[i]} P_{click[j]}}\label{eq:g2}\\
g_{[ijk]}^{(3)}&=&\frac{P_{click^{\otimes 3}[ijk]}}{P_{click[i]} P_{click[j]} P_{click[k]}}\label{eq:g3}\\
g_{[ijkl]}^{(4)}&=&\frac{P_{click^{\otimes 4}[ijkl]}}{P_{click[i]} P_{click[j]} P_{click[k]}P_{click[l]}}.
\label{eq:g4}
\end{eqnarray}


\begin{thebibliography}{99}
\bibitem{bondani}
M. Bondani, I. Degiovanni, M. Genovese, M. Paris and I. Ruo
Berchera, V. Schettini.  Found. of Phys. \textbf{41}, 305 (2011) AND REF.S THEREIN.
\bibitem{sramelow}
S. Ramelow  et al., Highly efficient heralding of entangled single photons," Opt. Express \textbf{21}, 6707-6717 (2013)
\bibitem{sorgentina}
G. Brida et al., Applied Phys. Lett. \textbf{101},  221112 (2012).
\bibitem{krapick}S. Krapick et al, An efficient integrated two-color source for heralded single photons, New J. Phys. \textbf{15},  033010 (2013)
\bibitem{fortsch}
M. F\"ortsch et al., A versatile source of single photons for quantum information processing, Nat. Comm.\textbf {4}, 1818 (2013)
\bibitem{montaut}
N. Montaut, High efficiency 'plug \& play'source of heralded single photons, arXiv:1701.04229
\bibitem{oxborrow}
M. Oxborrow, A. G. Sinclair, Single-photon sources, Contemp. Phys, \textbf{46}, 173-206 (2005)
\bibitem{eiseman}
M. D. Eisaman, J. Fan, A. Migdall, S. V. Polyakov,Invited review article: Single-photon sources and detectors, Rev. Sci. Inst. \textbf{82}, 071101 (2011)
\bibitem{sps}
C. J. Chunnilall, I. P. Degiovanni,S K\"{u}ck ,I.  M\"{u}ller I, A. G. Sinclair, Metrology of single-photon sources and detectors: a review. Opt. Eng. \textbf{53}(8), 081910 (2014) doi:10.1117/1.OE.53.8.081910.
\bibitem{dia1}
I. Aharonovic et al., Solid-state single-photon emitters, nat Phot \textbf{10}, 631 (2016)
\bibitem{dia2}
C. Kurtsiefer, S. Mayer, P. Zarda and H. Weinfurter, Stable Solid-State Source of Single Photons, Phys. Rev. Lett. \textbf{85} (2), 290 (2000).
\bibitem{dia3}
A. Beveratos, R. Brouri, T. Gacoin, J.-P. Poizat and P. Grangier, Nonclassical radiation from diamond nanocrystals, Phys. Rev. A \textbf{64}, 061802 (2001).
\bibitem{dia4}
A. Beveratos, S. Khn, R. Brouri, T. Gacoin, J.-P. Poizat and P. Grangier, Room temperature stable single-photon source, Eur. Phys. J. D \textbf{18},  191 (2002).
\bibitem{dia5}
D. Gatto Monticone et al.,Single-photon emitters based on NIR color
centers in diamond coupled with solid immersion lenses, Int. J. Quantum Inf. \textbf{12}, 1560011 (2014)
\bibitem{dia6}
T. Scr\"oder et al., Quantum nanophotonics in diamond,  JJ. Opt. Soc. Am. B 33, B65-B83 (2016)
\bibitem{grangier}
P. Grangier, G. Roger, and A. Aspect, Experimental evidence for a photon anticorrelation effect on a beam
splitter: a new light on single-photon interferences, Europhys. Lett.
\textbf{11}
, 173-179 (1986).
\bibitem{hbt}
R. Hanbury-Brown and R. Q. Twiss, Correlation between photons in two coherent beams of light, Nature (London) \textbf{177}, 27 (1956).
\bibitem{elizabeth} E. A. Goldschmidt et al., Phys. Rev. A \textbf{88}, 013822 (2013).
\bibitem{sres}
D. Gatto Monticone, Beating Abbe diffraction limit in confocal microscopy via non-classical photon statistics, Phys. Rev. Lett \textbf{113}, 143602 (2014).
\bibitem{vonz1}
A. Classen et al., Superresolving Imaging of Arbitrary One-Dimensional Arrays of Thermal Light Sources Using Multiphoton Interference, Phys. Rev. Lett. \textbf{117}, 253601 (2016).
\bibitem{vonz2}
S. Oppel, T. B\"uttner, P. Kok, and J. von Zanthier, Superresolving Multiphoton Interferences with Independent Light Sources, Phys. Rev. Lett. \textbf{109}, 233603 (2012).
\bibitem{Filip}
L. Lachman, L. Slodi\v{c}ka, \& R. Filip, Nonclassical light from a large number of independent single-photon emitters, Sci. Rep. \textbf{6}, 19760 (2016).

\bibitem{qudi}J. M. Binder, A. Stark, N. Tomek, J. Scheuer, F. Frank, K. D. Jahnke, C. Müller, S. Schmitt, M. H. Metsch, T. Unden, T. Gehring, A. Huck, U. L. Andersen, L. J. Rogers, F. Jelezko, Qudi: A modular python suite for experiment control and data processing, SoftwareX \textbf{6}  85 (2017). doi: 10.1016/j.softx.2017.02.001

\bibitem{Rondin} L. Rondin, G. Dantelle, A. Slablab, F. Grosshans, F. Treussart, P. Bergonzo, S. Perruchas, T. Gacoin, M. Chaigneau, H.-C. Chang, V. Jacques, J.-F. Roch, Surface-induced charge state conversion of nitrogen-vacancy defects in nanodiamonds, Phys. Rev. B \textbf{82}, 115449 (2010).

\bibitem{Pezzagna} S. Pezzagna, B. Naydenov, F. Jelezko, J. Wrachtrup, J. Meijer, Creation efficiency of nitrogen-vacancy centres in diamond, New J. Phys. \textbf{12}, 065017(2010).


\end{thebibliography}

\section*{Acknowledgements}

This research activity was supported by the following projects: EMPIR Project. No. 14IND05-MIQC2, Project Q-SecGroundSpace,  "DIESIS" project funded by the Italian National Institute of Nuclear Physics (INFN) - CSN5 within the "Young research grant" scheme. Ion irradiations were performed within the "Dia.Fab." experiment at the INFN-LNL laboratories, Italy.









\end{document}